\documentclass[RNAAS]{aastex62}

\usepackage{textcomp}
\usepackage{amsmath}

\begin{document}

\title{A new approximation of photon geodesics in Schwarzschild spacetime}

\correspondingauthor{Riccardo La Placa}
\email{riccardolaplaca81@gmail.com}

\author{Riccardo {La Placa}}
\affiliation{INAF - Osservatorio Astronomico di Roma, Monte Porzio Catone, Italy}
\affiliation{Research Centre for Computational Physics and Data Processing, Silesian University in Opava, Czech Republic}

\author{Pavel Bakala}
\affiliation{Research Centre for Computational Physics and Data Processing, Silesian University in Opava, Czech Republic}
\affiliation{M. R. \v{S}tef\'anik Observatory and Planetarium, Hlohovec, Slovak Republic}
\affiliation{INAF - Osservatorio Astronomico di Roma, Monte Porzio Catone, Italy}


\author[0000-0002-0018-1687]{Luigi Stella}
\affiliation{INAF - Osservatorio Astronomico di Roma, Monte Porzio Catone, Italy}

\author{Maurizio Falanga}
\affiliation{International Space Science Institute (ISSI), Bern, Switzerland}
\affiliation{International Space Science Institute Beijing, P.R. China}


\section{} 

Photon geodesics in the Schwarzschild metric lie on the plane defined by the central mass in the origin, the emitting point and the observer. The elliptic integral which describes them has often been approximated analytically in terms of the impact parameter $b$, a conserved quantity which in practical (numerical) applications often requires \textit{a priori} knowledge of the trajectory's characteristics \citep[see \textit{e.g.}][]{Semerak2015}. 
\citet{Beloborodov2002} proposed the following approximation (later derived in \citet{DeF2016} by Taylor-expanding the geodesics equation)
\begin{equation}\label{eq:bel}
	1 - \cos \alpha \approx (1 - \cos \psi)(1 - 2r_{g}/r) \ , 
\end{equation}
where $r_{g} = GM/c^{2}$ is the gravitational radius. Knowing the emission radius $r$ and the angle $\psi$ between the line of sight and the radial direction $\hat{r}$, Eq. \ref{eq:bel} gives the emission angle $\alpha$ between $\hat{r}$ and the initial direction needed for the photon to reach the observer at infinity parallel to the line of sight with $ b = r (1 - 2r_{g}/r)^{-\frac{1}{2}} \sin \alpha  $ \citep[see Fig. 1 in][]{Beloborodov2002}. This is an especially convenient set of variables for applications in which fast calculations of photon geodesics from the emitting point to the observer are required \citep[see \textit{e.g.}][]{Chang2006, Poutanen2006}. Eq. \ref{eq:bel} yields a small deviation from the exact trajectories ($\delta\alpha/\alpha < 1$\%) for \textit{direct} photons ($\alpha \leqslant \pi/2 $); its accuracy, however, degrades from $\delta\alpha/\alpha \sim 1$\% up to $10$\% for a wide range of geodesics with a turning point ($\alpha > \pi/2 $).

\begin{figure}[h!]
\begin{center}
\gridline{\fig{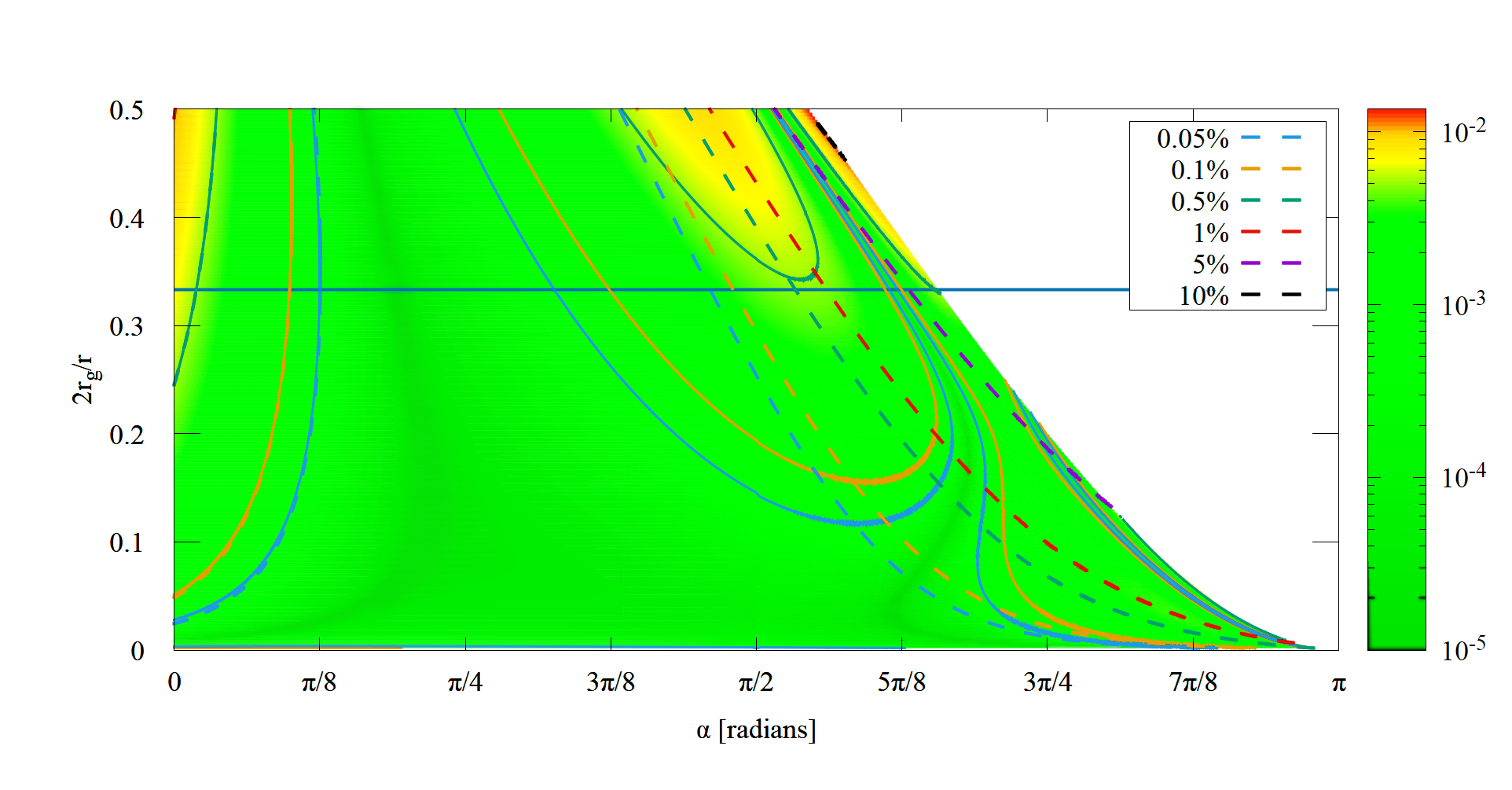}{1.0\textwidth}{}}
\vspace{-1.2cm}
\gridline{
	\fig{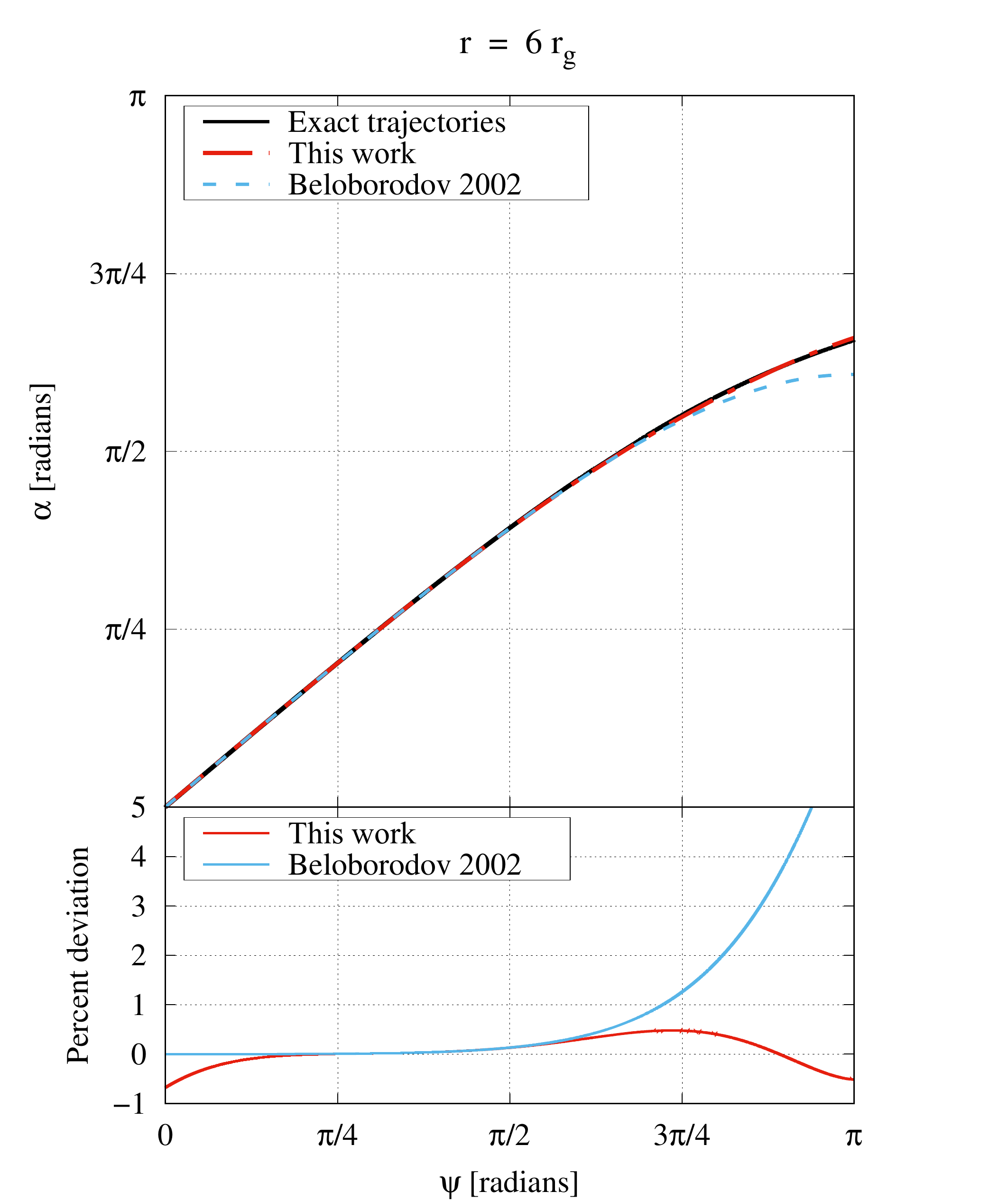}{0.49\textwidth}{}
	\fig{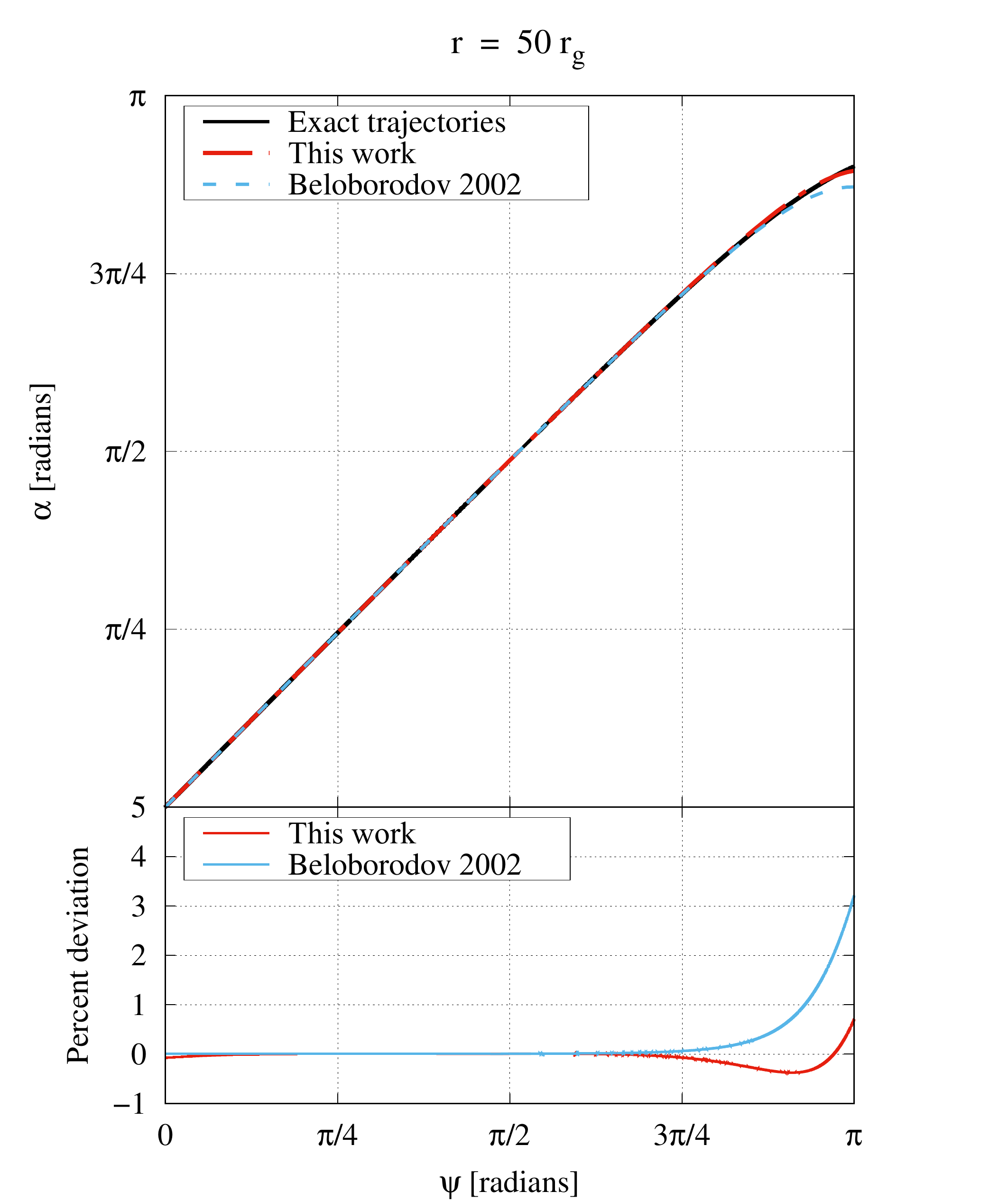}{0.49\textwidth}{}
}
\caption{\\ \textit{Top -} Color-coded accuracy $ \delta\alpha/\alpha$ of Eq. \ref{eq:new} as a function of $r_{g}/r $ and $\alpha$ (for $0 \leqslant \alpha \leqslant \alpha(r, \psi = \pi)$). The color scale was cut at $ \delta\alpha/\alpha = 10^{-5} $ for clarity. The horizontal blue line represents $ r_{ISCO}= 6~r_{g} $. Overplotted are selected contour levels of $\delta\alpha/\alpha = const$ both for Eq. \ref{eq:new} (solid lines) and Eq. \ref{eq:bel} (dashed lines), with equal colors representing equal inaccuracies (to avoid cluttering, contour levels below $0.05$\% are not plotted). \\ \textit{Bottom -} The two upper panels show the behavior of Eq. \ref{eq:bel} and \ref{eq:new} for $0 \leqslant \psi \leqslant \pi$ at two sample emission radii, $ r = r_{ISCO}$ and $ r = 50~r_{g}$; the lower panels show the deviation of each approximation from the numerical results.}
\end{center}\label{fig:1}
\end{figure}

We present the following approximate equation which holds for $\psi$ between 0 and $\pi$ (and thus up to the maximum $\alpha$ we consider, $ \alpha(r, \psi = \pi) )$ and retains $\delta\alpha/\alpha \lesssim 1$\% throughout:
\begin{equation}\label{eq:new}
1 - \cos \alpha \approx (1 - \cos \psi)(1 - \dfrac{2r_{g}}{r})[1 + k_{1} \dfrac{2r_{g}}{r} (1 - \cos(\psi-k_{2}))^{k_{3}}]\ ;
\end{equation}
here $k_{1} = 0.1416,~k_{2} = 1.196,~k_{3} = 2.726$. 
The form of Eq. \ref{eq:new} and the values of the $k$-parameters were determined by fitting the relationship between $ \alpha $ and $ \psi $ to the geodesics computed numerically for various radii between 6 and 100~$ r_g $, and by tuning the parameters for $r = 10~r_g$. Fig. \ref{fig:1} shows the accuracy of Eq. \ref{eq:new}: for radii larger than $r_{ISCO} = 6~r_g$ (the radius of the innermost stable circular orbit for massive particles) our approximation yields $\delta\alpha/\alpha \lesssim 0.7\% $ over all angles. For lower radii $\delta\alpha/\alpha$ crosses the $1\%$ threshold in a small region with $r \lesssim 4.7~r_g$, reaching a maximum of $1.35\%$ for a radius of $4~r_g$, below which the approximation breaks down.
Also note that by using Eq. \ref{eq:new} while calculating $ \partial b/\partial r $ in the expression of the solid angle element \citep[see \textit{e.g.}][]{Bao1992}, a higher accuracy is attained than applying the approximation derived in \citet{DeF2016}.
Eq. \ref{eq:new} is especially useful in geometries for which $\psi$ approaches $\pi$, where Eq. \ref{eq:bel} would produce overly bent trajectories \citep[see \textit{e.g.} the application to the ``far side'' of a relativistic accretion disk in][]{mainart}.


%


\begin{thebibliography}{}

\bibitem[Bao(1992)]{Bao1992} Bao, G.\ 1992, \aap, 257, 594
\bibitem[Beloborodov(2002)]{Beloborodov2002} Beloborodov, A.~M.\ 2002, \apj, 566, L85
\bibitem[Chang et al.(2006)]{Chang2006} Chang, P., Morsink, S., Bildsten, L., et al.\ 2006, \apj, 636, L117
\bibitem[De Falco et al.(2016)]{DeF2016} De Falco, V., Falanga, M., \& Stella, L.\ 2016, \aap, 595, A38
\bibitem[La Placa et al.(2019)]{mainart} {La Placa}, R., Stella, L., Papitto, A., et al. \ 2019, In preparation
\bibitem[Poutanen \& Beloborodov(2006)]{Poutanen2006} Poutanen, J., \& Beloborodov, A.~M.\ 2006, \mnras, 373, 836
\bibitem[Semer{\'a}k(2015)]{Semerak2015} Semer{\'a}k, O.\ 2015, \apj, 800, 77

\end{thebibliography}
\end{document}